# Analysis of historical meteor and meteor shower records: Korea, China, and Japan

Hong-Jin Yang [a, b,*], Changbom Park [c], and Myeong-Gu Park [b]

[a] Korea Astronomy Observatory [†], Daejeon 305-348, S. Korea.
[b] Department of Astronomy and Atmospheric Sciences, Kyungpook National University, Daegu 702-701, S. Korea.
[c] School of Physics, Korea Institute for Advanced Study, Seoul 130-722, S. Korea.

------------------------------------------------------------------------------------------------

**Abstract**

We have compiled and analyzed historical Korean meteor and meteor shower records in three Korean official history books, *Samguksagi* which covers the three Kingdoms period (57 B.C~A.D. 935), *Goryeosa* of Goryeo dynasty (A.D. 918~1392), and *Joseonwangjosillok* of Joseon dynasty (A.D. 1392~1910). We have found 3861 meteor and 31 meteor shower records. We have confirmed the peaks of Perseids and an excess due to the mixture of Orionids, north-Taurids, or Leonids through the Monte-Carlo test. The peaks persist from the period of Goryeo dynasty to that of Joseon dynasty, for almost one thousand years. Korean records show a decrease of Perseids activity and an increase of Orionids/north-Taurids/Leonids activity. We have also analyzed seasonal variation of sporadic meteors from Korean records. We confirm the seasonal variation of sporadic meteors from the records of Joseon dynasty with the maximum number of events being roughly 1.7 times the minimum. The Korean records are compared with Chinese and Japanese records for the same periods. Major features in Chinese meteor shower records are quite consistent with those of Korean records, particularly for the last millennium. Japanese records also show Perseids feature and Orionids/north-Taurids/Leonids feature, although they are less prominent compared to those of Korean or Chinese records.

---

[*] Corresponding author. Fax:+ 82-42-861-5610
  E-mail address: hjyang@kao.re.kr, hjyang@kasi.re.kr (H.-J. Yang)
[†] It has changed the name for 'Korea Astronomy and Space Science Institute' as from Jan. 1, 2004

# I. Introduction

## 1.1 Korean records

The relationship between meteor shower and comet was known in the 18th century. Meteor showers that display bright streaks of light in the sky are caused by meteor streams. The meteor streams are the debris of comets. They are spread out in time by collisions, planetary perturbation, and solar radiative pressure. Particles gradually become smaller until we cannot see meteor shower events any more. The timescale of this is known to be about 3~7 years for Leonids (Yeomans 1981; Williams 1997; Wu 2002).

While meteors vary in size and velocity, naked-eye observers can see only the meteor events brighter than about 5th magnitude. Most of meteors burn up in the atmosphere, but some can reach the surface of the Earth, which we call meteorites. When large number of meteors fall in a relatively short time and in approximately parallel trajectories, we call it a meteor shower. Meteor shower is one of the most spectacular astronomical events in the night sky. Naturally, many records of them appear in the historical documents of countries like Korea, China and Japan as well as European and Arab countries, all of which had a keen interest in astronomical phenomena.

Korean history books contain records of numerous astronomical phenomena, such as meteor, meteor shower, and meteorite as well as solar and lunar eclipse, planetary event, sunspot, aurora and so forth. Among diverse history books, there are so called "official history books (正史)" which are written under the auspice of the court. The unique value of Korean official history book is that astronomical phenomena were systematically recorded by the court astronomers and compiled into a few most official history books by dynasties that lasted for from 500 to 1000 years. Interesting thing is that during Joseon dynasty (A.D. 1392~A.D. 1910) the political system has been set in such a way that recording of history is kept strictly unbiased, and even kings were not in general allowed to read these history books about his own reign nor about his predecessors' (cha 1983). Another characteristic of Korean history is that each dynasty has lasted at least for about 500 years unlike Chinese or Japanese dynasty. Because of longevity of each dynasty, Korean history during the last 2000 years, from 57 B.C. to A.D. 1910, is recorded in only three official history books:

*Samguksagi**\** (三國史記, the History of the Three Kingdoms; Kim et al., 1145), *Goryeosa* (高麗史, the History of the Goryeo dynasty; Kim et al., 1451), and *Joseonwangjosillok* (朝鮮王朝實錄, the Annals of the Joseon dynasty; Chunchugwan, 1392-1863).

*Samguksagi* covers the period from 57 B.C. to A.D. 935, the period of Three Kingdoms (三國), namely Silla (新羅), Goguryeo (高句麗), and Baekjae (百濟). It contains total of 236 records of solar eclipses, comets, meteors, planet motions, etc. The first astronomical record in *Samguksagi* is a solar eclipse event in 54 B.C. (the 4th year of Hyeokgeose [赫居世]), which is confirmed to have actually occurred (Park & La 1994). The whole astronomical records from *Samguksagi* have been studied by Park & La (1994). They have proved that the astronomical records in *Samguksagi* are those that cannot be formed by simply copying Chinese records, particularly in eclipse records. *Samguksagi* has 39 meteor events including 13 meteor shower events. Only 5 meteor records have precise date information while the remaining ones have only the year and month information. The first record of meteor shower appears in A.D. 104 (the 25th year of Pasa [婆娑]). The majority of the astronomical records in this period are described in a few words. Likewise, half of the meteor shower events are also described simply as "Stars fell down like rain (星隕如雨)". The rest are described like "Many stars flew in a certain (or all) directions in the sky".

*Goryeosa* covers the period of Goryeo dynasty (高麗) from A.D. 918 to A.D. 1392. It is a well-arranged history book divided into several chapters by contents, and records are listed in chronological order. The most abundant astronomical records in *Goryeosa* are the planetary motions (~ 1300 records) and the meteor events are the second most abundant ones, followed by aurora and sunspot records (Yang et al. 1998). *Goryeosa* contains 740 records of meteor events including mere 9 meteor shower events. Most of them have positional information like constellation or direction. A few bright meteors have additional information such as size and motion of the trail. All meteor showers were described as flows of many stars. Although a meteor shower displays more dramatic scene than an ordinary meteor, in *Goryeosa* meteor events were recorded in more detail than meteor showers. For instance, while a meteor shower is described simply as "Many small stars flew in all directions (衆小星分流四方)", the record on a meteor reads "A meteor with the size of a pear and with 1.8 meter trail flews from Urim (羽林: one of the oriental constellations) into Wi (危: another oriental constellation)".

---

\* All Korean historical names in this paper follow the notation used by the Korean government.

The most recent history book, *Joseonwangjosillok* (*Sillok*, hereafter), covers Joseon dynasty (朝鮮) from A.D. 1392 to A.D. 1910. It has 3111 records of meteor events including 7 meteor shower events. All records in Sillok have vivid and descriptive accounts on position, direction, size, and color of meteors like those in *Goryeosa*. The entire Sillok is now available on CD-ROM.

One record in 1490 (the 21st year of King Seongjong [成宗]) reads "Since we can see meteors everyday, we shall record only the special meteors". In spite of this instruction, we could not find any evidence that the number of meteor records had actually diminished since then, except for the following few years. Meanwhile, all 7 meteor shower records in Sillok contain the description of the size, color, and direction of the trails.

The records on meteor events occupy a large percentage of the historical astronomical records of Korea. The fraction of meteor records is nearly constant from the Three Kingdoms period to Joseon dynasty. For example, 16.5% of all 236 astronomical records in *Samguksagi* are meteor records, 14.7% of about 5000 astronomical records in *Goryeosa*, and 16.7% of about 18700 astronomical records in *Sillok* are meteor events. Meteorite is not a frequent astronomical event, but Korea has 54 meteorite records in the three official history books. The numbers of meteor, meteor shower, and meteorite observations during each dynasty are listed in Table 1. The compilation and analysis of all astronomical records from these official history books will be given elsewhere (Park & Yang 2004, in preparation).

**(Table 1)**

1.2 Chinese and Japanese records

China and Japan also have a long history of astronomical observations. Chinese astronomical records covering a longer period have been compiled in *General Compilation of Chinese Ancient Astronomical Records* (中國古代天象記錄總集; Beijing Observatory 1988). The book lists historical Chinese records up to A.D. 1911 of sunspot, aurora, meteor, meteorite, solar eclipse, lunar eclipse, comet and so forth from a number of history books. The data are conveniently arranged in chronological order for a given class of event. The total number of meteor records is about 5700, and the number of meteor shower records, as classified in the book, is 276.

Historical Japanese astronomical records are compiled in three books, *Japanese Historical Astronomical Records* (日本天文史料; Kanda 1935), *General Inventory of Japanese Historical Astronomical Records* (日本天文史料綜覽; Kanda 1936), and *Japanese Historical Astronomical Records after 1600* (近世日本天文史料; Ohsaki 1994). The first and second ones cover the period before 1600 and the last one covers the period after 1600. *Japanese Historical Astronomical Records* and *General Inventory of Japanese Historical Astronomical Records* have the same contents but are in different formats. They list 431 records of meteors and 13 records of meteor showers.

## 1.3 Previous works on Korean records

Korean meteor and meteor shower records have been studied in conjunction with Chinese and Japanese records. Imoto & Hasegawa (1958) published historical meteor shower records from China, Korea, and Japan. They compiled 118 meteor shower records; 59 from China, 36 from Korea, and 23 from Japan. While Chinese and Japanese meteor shower records were compiled from 19 and 17 historical books, respectively, Korean records were compiled from only two books, *Yeoulseongsillok* (列聖實錄) and *Munheonbigo* (文獻備考), which are the rearranged records compiled from the official Korean history books. Unfortunately, they sometimes are not as accurate as the original sources, particularly in dates.

Hasegawa (1998) studied Korean meteor record again, but his work was limited to Goryeo dynasty. Mason (1995) compiled and identified world-wide historical Leonids records. Mason listed 58 records related to outstanding showers and meteor storms of Leonids including 6 Korean records from Imoto & Hasegawa (1958). Meanwhile, Hasegawa (1980, 1992) studied Chinese and Japanese meteor shower records. Rada & Stephenson (1992) report 26 meteor shower events from the near 10th century Arabic and the medieval European records.

Recently, Ahn et al. (2002) and Ahn (2003) analyzed Korean meteor and meteor shower records during Goryeo dynasty. They have identified peaks in the distributions of meteor records with Perseids, Leonids, eta-Aquarids, and Orionids. Seasonal variation of sporadic meteors was also qualitatively discussed. In their works, however, the time between two successive perihelion passages of the Earth, i.e. anomalistic year, has been erroneously used as the basic period of showers.

In this work, we have compiled the meteor and meteor shower records directly from all three official Korean history books, *Samguksagi*, *Goryeosa*, and *Sillok*. We analyze these records to study the long-term behaviors of meteor shower phenomena and related comets. We also compare our Korean records with Chinese and Japanese ones. We then discuss the implication of our results on the modern meteor showers and comets. Additionally, we discuss seasonal variation of sporadic meteors based on the abundant meteor records of Goryeo and Joseon dynasties.

## II. Analysis of Korean records

2.1 Korean meteor and meteor shower records

All historical Korean astronomical observations are recorded in luni-solar calendar. We have converted all dates into solar calendar dates and Julian days by using *Korean Chronological Tables* (Shim et al., 1999; Ahn et al., 2000a, Ahn et al., 2000b; Han, 2001). Many records during the Three Kingdoms period have only the year and month information. Records of Goryeo and Joseon dynasties have specific date information, and the conversion is straightforward. A very small fraction of records in these periods has erroneous or ambiguous date information.

Meteor shower is caused by the meteor stream which is the debris of a comet. If the stream remains at the same position in space, the radiant of meteor shower appears at the same constellation as the Earth repeatedly passes through this stream during its yearly revolution around the sun. The period of this recurring shower will be one sidereal year (365.2564 days) if the orbit of the meteor stream and the orbit of the comet do not change over the years. We use this sidereal year as the basic period for yearly repetition of meteor shower. Although its differences from the tropical year (365.2422 days) and the anomalistic year (365.2596 days) are small (0.0142 and 0.0032 days per year, respectively), they can accumulate to significant amount when historical astronomical records over a thousand years are considered.

We plot in Fig. 1 all meteor (cross) and meteor shower (circle) events recorded in official Korean history books. Meteor records are much more numerous than meteor shower records. We mainly use meteor records for statistical analysis. To confirm the existence of periodic meteor showers, we have calculated the power spectrum of all meteor record events in time using the one-dimensional Fourier transform, and found a dominant peak in the power spectrum at one year period.

**(Figure 1)**

The histogram in Fig. 2 shows the accumulated number for each period of meteor records in 5 day bin as a function of the "counting day" within each sidereal year. January 1, 2000 is set to be the day 1 of the reference sidereal year. The counting day for each meteor record can be expressed as follows;

Counting Day = Julian day of each record –
$$\text{(Julian day of Jan. 1, 2000} -n*\text{sidereal year),} \qquad (1)$$

where n is a positive integer that makes the counting day a value between 0 and 365.2564.

**(Figure 2)**

The upper panel (Fig. 2a) is for the Three Kingdoms period, the middle one for Goryeo Dynasty, and the lower one for Joseon Dynasty. Open circles in the upper part of each panel show meteor shower records. Horizontal error-bars and reversed triangles above the box indicate durations and the maximum dates of the well-known meteor showers as of A.D. 2000. Bold characters P's above the reversed triangles represent modern periodic meteor showers with known associated comets.

To determine if these are statistical noise or not we use a Monte Carlo test. Since the observation of meteors is affected by seasonal conditions, Monte Carlo samples are randomly generated from the distribution of general astronomical records compiled from the official Korean history books (Park & Yang 2004, in preparation). They are the records of various astronomical phenomena with no intrinsic seasonal variation. Any seasonal variation in these records must be due to the observational bias. For the Three Kingdom period, 179 records of various astronomical observations are selected out of total 236 to construct the parent distribution for Monte Carlo samples. Records of planetary motions are the most abundant during Goryeo and Joseon dynasties, and they are expected to reflect the observational selection effects of astronomical records in history books. For Goryeo dynasty, 3534 planetary and lunar records are selected out of total about 5000 astronomical records. For Joseon dynasty, 18697 planetary records out of the total ~187000 are used to represent the general distribution of astronomical records. As is expected from the conditions of astronomical observation in Korea due mainly to seasonal weather, the distribution of planetary records shows an excess from September to March relative to from May to August (Fig. 3).

**(Figure 3)**

Monte Carlo samples were randomly generated from the distribution of these selected records over the period of sidereal year. Ten thousand sets of mock samples, each having the same total number of records as the actual meteor and meteor shower records, were produced, and at each day bin, the numbers of generated records below which 90% (1.7 σ) or 99.9% (3 σ) of the total Monte Carlo samples are located are determined. The dot-dashed line in Fig 2 represents the 90% level, and the solid line 99.9%. For the Three Kingdoms period (Fig. 2a), only one small peak near day 195 is above 90% line, and none is above 99.9%. In this period, there are only a few meteor records despite the long historical span of roughly 1000 years, and most of them do not have precise date information. So we do not try to identify these peaks with any present meteor showers.

Most of meteor and meteor shower records during the Three Kingdoms period (Fig. 2a) give only the year and month of the events but no date. We, therefore, smooth all records over the corresponding lunar month. One record appearing in a certain lunar month contributes 1/N event to each day over N days of that lunar month. Probably due to the small number of total records and lack of date information, it is difficult to say whether the distribution of records in Fig. 2a shows any obvious correlation with modern meteor showers.

The histogram and open circles in Fig. 2b show the distributions of meteor and meteor shower records during Goryeo dynasty. Unlike the three Kingdoms period, a very prominent peak appears at the day 220. Three of the eight meteor shower records of Goryeo dynasty are also located at nearly the same day. It coincides with the peak day of current Perseids, which is around August 12. We identify this peak with Perseids whose parent comet is known to be 109P/Swift-Tuttle.

The next notable feature is the broad heap of peaks at days 270~340. The peaks generally coincide with current Leonids, Orionids and north-Taurids, whose durations often overlap with one another. It is, therefore, not straightforward to separate and associate each peak to one of the three showers.

The histogram and open circles in Fig. 2c show meteor and meteor showers during Joseon dynasty. Both peaks at the day 220 and the broad bump at days 270 ~ 340 are conspicuous. The position of the peak, day 220, is exactly the same as that during the Goryeo dynasty, but the relative height (i.e., number of records) of the peak above the background is lower compared to that of Goryeo dynasty. We again identify this peak with Perseids. The same location of the peak implies the orbit of its parent comet 109P/Swift-Tuttle has been stable over the last 1000 years without significant perturbation on it. However, the decrease of the ratio of Perseids records to the background records from Goryeo dynasty to Joseon dynasty suggests the activity of Perseids has been decreasing over the last 1000 years.

The shape of the broad bump at days 270~340 during Joseon dynasty is somewhat different from that during Goryeo dynasty. It has two slightly separated peaks. The first one is in accord with Orionids. The second one spans the period from the day 305 to 325 (1st~21st Nov.), which coincides with north-Taurids and Leonids. Comparison of Fig. 2b with Fig. 2c indicates that the broad bump between the day 270 and 340 is more prominent during Joseon dynasty than during Goryeo dynasty. We identify the bump of peaks around the day 290 with Orionids and that around day 310 with north-Taurids and Leonids.

Seven records of meteor showers during Joseon dynasty are also plotted in Fig. 2c as open circles with sizes proportional to the number of records. Two shower records at the day 250 do not explicitly coincide with neither the peak of meteor records nor present meteor showers. But the four shower records at the day 315 and the one at the day 320 seem to coincide with the peak of meteor records as well as present north-Taurids/Leonids.

From the Monte Carlo test it is known that peaks at the day 220 (Perseids) during Goryeo (Fig. 2b) and Joseon dynasties (Fig. 2c) are well above the 99.9% level as expected. Also, a broad bump at days 270~340 during Goryeo dynasty is above the 90% level whereas the similar bump at days 285~325 during Joseon dynasty is even above the 99.9% level. Hence, existence and activity of the Perseids as well as Orionids, north-Taurids/Leonids during the last 1000 years are evident with high statistical significance. We may conclude that the persistent existence of a peak at the day 220 is a proof that Perseids has been active at least for 1000 years, and also that the orbit of the parent comet of Perseids, 109P/Swift-Tuttle, has been similar to the current one during the last 1000 years.

There are also significant peaks at days 265, 310 and 335 during Goryeo dynasty and at days 290, 315 and 335 during Joseon dynasty at 99.9% level. While the peak at the day 335 remains at the 99.9% level throughout Goryeo and Joseon dynasties, the peaks near the days 290 (October 17) and 315 (November 11) strengthen from Goryeo to Joseon dynasty. This implies that the activities of Orionids and north-Taurids/Leonids have been increasing over the last 1000 years. Hence, their parent comets, Halley for Orionids and Encke for north-Taurids, have been increasingly active in stable orbits.

Now, we move on to discuss the "explicit" meteor shower records in Korean history books. Although meteor shower records are not abundant, Korean history books contain well-described meteor shower records. We list in Table 2 all Korean historical meteor shower records taken from the three official history books and *Jeungbomunheonbigo* (增補文獻備考; Chanjipcheong, 1908). *Jeungbo-munheonbigo* is an encyclopedia published during the late Joseon dynasty, which includes the compilation of astronomical records in all previous periods. Most of these meteor shower records are listed in Imoto & Hasegawa (1985), but we find five new records not listed in Imoto & Hasegawa (1958). These records are marked as "3" in the last column of Table 2. On the other hand, they have listed additional 10 meteor shower records. One record of the year 1410 seems to be about atmospheric turbulence effect. Most of the rest are simple meteor events.

<p align="center">**(Table 2)**</p>

Therefore, we confirm 31 records listed in Table 2 as the meteor shower records documented in the official Korean history books and *Jeungbomunheonbigo*. We have identified 11 among the 31 Korean historical records as periodic meteor shower records. The counting day, which is defined in Eq. (1), of Korean records is compared with that of current periodic meteor showers. Using this method, we find three Leonids and one eta-Aquarids records. However, for Leonids, we refer Mason (1995) because Leonids does not appear at the same counting day. According to Mason (1995), Newton (1864) found that Leonids appeared delayed at a rate of 1.45 day per century relative to sidereal year because the orbit of parent comet 55P/Tempel-Tuttle is perturbed by Jupiter, Saturn and Uranus. From Mason' estimated dates of Leonids, we confirm 4 Leonids records again and find two new candidate records of Leonids in 801 and 1111. The record of 801 does not have day information. However, its converted month overlaps the peak of Leonids. We confirm the record of the year 1111 as a Leonids by interpolating tabulated dates from Mason (1995).

The precession of the orbit of comet 55P/Tempel-Tuttle makes the peak of the Leonids appears between days 312 and 320 during Joseon dynasty, and between days 305 and 312 during Goryeo dynasty, and mainly between days 299 and 305 during the Three Kingdoms period. In Fig. 2b and 2c, we can find local maxima at days corresponding to these peak intervals. They are to be compared with current Leonids peak at the day 321. The delayed appearance of Leonids may also contribute to the broad bump around days 305~320 in both Fig. 2b and 2c. Therefore, Korean records are consistent with the nodal advance of Leonids, and reveal the orbital perturbations of 55P/Tempel-Tuttle by planets.

2.2 Seasonal Variation

According to Yrjölä & Jenniskens (1998), activity of seasonal variation of sporadic meteors is expected to obey the following formula,

$$N(H,\lambda_\odot) = <N_{spo}>(H) - \Delta N_{spo}(H) \cos(\lambda_\odot), \qquad\qquad (2)$$

where $<N_{spo}>$ is the mean daily sporadic hourly count at the summer solstice (June 21), $\Delta N_{spo}$ is the yearly amplitude due to the seasonal variation at a given local time of the day (H), and $\lambda_\odot$ is the solar longitude measured from the Vernal equinox. On the northern hemisphere, rates are expected to peak at the Autumnal equinox (about September 21) and to be lowest at the Vernal equinox (March 21). Since the seasonal variation of sporadic meteors is caused by revolution of the Earth around the Sun, the periodic activities should be repeated over the tropical year.

Korean meteor records do not contain hourly counts. Instead, we have a large number of meteor records for a given day, accumulated over hundreds of years. These records are undoubtedly expected to reflect seasonal variations. We, therefore, modify Eq. 2 to

$$N(\lambda_\odot) = <N_{spo}> - \Delta N_{spo} \cos(\lambda_\odot + \Phi) \qquad\qquad (3)$$

to describe the accumulated number distribution of meteor records. The phase $\Phi$ accounts for the possible phase shift.

Historical Korean records were affected by seasonal variations of different origins, particularly because the observational condition in Korea strongly depends on the season. We expect that the seasonal distribution of the general astronomical records, estimated from records on planetary and lunar motions, faithfully reflects the seasonal condition for astronomical observations (Fig. 3). We call this background distribution "the seasonal observational bias". It varies over seasons mainly because astronomical observations are influenced by the weather condition. The seasonal observational biases from Goryeo dynasty (Fig. 3a) and Joseon dynasty (Fig. 3b) show a good similarity. However, the seasonal variation of meteor records (see Fig. 2) is noticeably different from the seasonal observational bias. Hence, the distribution of meteor records seems to contain inherent seasonal variation in addition to the variation induced by the seasonal observational bias.

We analyze all meteor records from *Goryeosa* and *Sillok* to find evidence for the seasonal variation of sporadic meteors. To reduce contamination in the meteor records from specific meteor showers, the bins corresponding to the peak at day 220 and to the bump at days 285~325 have been excluded in fitting the seasonal variation curve. This pruned number distribution is divided by the normalized seasonal observational bias to produce the inherent seasonal variations of sporadic meteors for Goryeo and Joseon dynasties, which are shown in Fig. 4 by open circles. The number of records is systematically higher at the latter half of the year in both periods. The seasonal variation is very weak in Fig. 4a, but Fig. 4b clearly shows a seasonal variation.

<div align="center">(Figure 4)</div>

We used Eq. 3 to fit the inherent number distribution by the nonlinear least-square fitting method. The solid curve in Fig. 4b shows the best fit with $<N_{spo}>$ = 110.8±3.9, $\Delta N_{spo}$ = 27.3±5.6, and $\Phi$ = 15±11 days. The dotted curve represents Eq. 2 with the same amplitude. We therefore estimate that, on average, $\Delta N_{spo}/<N_{spo}> \approx$ 1/4.1, i.e., the maximum number of sporadic meteors within a year is roughly 1.7 times the minimum.

Hasegawa (1992) has shown the monthly variation of Chinese meteor records during A.D. 1~1900 using the tropical year as the base period. Monthly variation of Chinese meteor records is not exactly in accord with Korean one. Hasegawa noted that Chinese records have two maxima in July-August and October-November, and pointed out that the maximum in November might be related to the north-Taurids. He plotted monthly variations of meteors for each century after A.D. 1001. Both maxima tend to be progressively delayed: in the 11th and 15th centuries, maxima appear at July and October whereas in 18th and 19th centuries they appear at August and November. Since the difference between the sidereal year and the tropical year amounts to approximately 1.4 days per century, we expect to see the position of the maxima shifted by approximately half a month over 1000 years if records are arranged in tropical year base. Hence we believe the data are consistent with the proposition that the maxima shown in Hasegawa (1992) are Perseids and north-Taurids, and the change of maximum position is because the records are arranged in the tropical year.

## III. Chinese and Japanese records

### 3.1 Chinese meteor shower records

Chinese history comprises many dynasties of different origins, and often more than one dynasty have existed at a given time. As a result, astronomical records are scattered in many history books that were written by different dynasties under different circumstances. These massive yet uneven records are collected in *General Compilation of Chinese Ancient Astronomical Records* (中國古代天象記錄總集; Beijing Observatory 1988), and we use this book as the source of Chinese meteor shower records. Dates of all records are already converted into solar calendar in this book.

The meteor shower records are separately classified in this book. There are enough records to study the periodic meteor shower events. Zhang (1977) has reported that 69 of 103 Chinese meteor shower records are related to current meteor showers. However, Chinese meteor shower records have not been analyzed by statistical methods yet.

To facilitate comparison with Korean data we divide all Chinese meteor shower records into three sets corresponding to the same periods: (I) before 918 (Fig. 5a), (II) from 918 to 1392 (Fig. 5b), and (III) from 1393 to 1911 (Fig. 5c). Unlike Korean meteor records, where a significant number of records appear in the period (II) (Fig. 2b), most Chinese meteor shower records are concentrated in the period (III).

**(Figure 5)**

In the period (I), two peaks, one around day 120 and the other around day 220, are noticeable (Fig. 5a). Although the former coincides with the current eta-Aquarids, the debris of Halley, it may not be associated with eta-Aquarids because the peak weakens in (II), and finally, disappears in (III). So it may be attributed to some other ancient periodic meteor shower that has been active only up to the the 10th century, or to overlapping of random meteor showers.

The latter peak around the day 225 is conspicuous throughout the whole history although it is not so dominant during the period (III). We can safely identify this peak as Perseids. Also interesting is that this peak weakens or smears from the period (II) to (III) relative to the background or other features. This is also true for Korean records although the change is less prominent. This may suggest that the actual activity of Perseids has weakened over the last 600 years or so. Another rather broad bump with one noticeable peak exists around days 280~340. It agrees well with the bump around days 270~340 seen in the Korean data. But the details of the peak features are different. Only one outstanding peak appears near the day 315 (Fig. 5c). A peak at the same location also exits in Korean records (Fig. 2c) although it is somewhat wider. In Korean records, another peak stands out within the bump near the day 290 (Fig. 2c), which is absent in Chinese records (Fig. 5c).

3.2 Japanese meteor and meteor shower records

Since Japan also has only a limited number of meteor shower records, we analyze the meteor records instead. The 444 meteor and meteor shower records collected from various Japanese history books (Kanda 1935, Ohsaki 1994) are similarly divided into three periods as were done for Korean and Chinese data: (I) before 918, (II) from 918 to 1392 , and (III) from 1393 to 1867. We adopt the date conversion as appeared in *Japanese Historical Astronomical Records* (Kanda 1935) and *Japanese Historical Astronomical Records after 1600* (Ohsaki 1994).

**(Figure 6)**

There are roughly twice more Japanese meteor records in the period (I) compared to Korean ones (Fig. 6a). One rather dominant peak appears at the day 20, which is absent in both Korean meteor and Chinese meteor shower records at all periods. However, this peak disappears in the periods (II) and (III) in the Japanese records, and no modern periodic meteor shower is shown near the day 20.

Two noticeable bumps, one between the days 215 and 230 and the other between the days 290 and 320, appear during the period (II) (Fig. 6b). The positions of peaks are consistent with those in Korean and Chinese records for the same period (see Fig. 2b and 5b). Hence, the former can be associated with Perseids, and the latter with Orionids, north-Taurids, and Leonids.

While China has more than 200 meteor shower records and, Korea has more than 3000 meteor records in the period (III), Japan has only 283 meteor records. One significant peak appears near the day 310, which can be associated with Orionids, north-Taurids, and Leonids as in Korean and Chinese records (Fig. 6c). However, there is no notable feature that can be associated with Perseids (Fig. 6c), which clearly manifests its presence in Korean and Chinese records during the last 1000 years (Fig. 2c and 5c).

Japanese meteor records seem to be least in common with others among Korean, Chinese, and Japanese meteor data, and have weak consistency among different periods. Although Korean records during the Three Kingdoms period do not show any prominent periodic meteor shower features, Japanese records before the year of 918 shows a prominent peak near day 20. Moreover, the peak dose not seems to share any significant common features with Korean or Chinese records, nor with those of Japanese records in the later periods. Analysis of Japanese solar eclipse records (see, Fig. 1 in Park 1996) also shows the records before the 10th century is in poor agreement with actual eclipse events.

We also note that the two dominant identified peaks, one peak at the day 215 in the period (II) and at the day 315 in (III), are shifted by one bin (5 days) compared with those in Korean and in Chinese records. This shift of peaks may be due to systematic error in the Japanese compilation of astronomical records.

## IV. Summary and Discussion

We have compiled and analyzed the meteor and meteor shower records in official Korean history books dating from 57 B.C. to A.D. 1910, covering the Three Kingdoms period, Goryeo dynasty, and Joseon dynasty. The books contain only a small number of meteor shower records in contrast to abundant meteor records.

The earliest records during the Three Kingdoms period are too few and have mostly year and month information only. However, the meteor records during Goryeo and Joseon dynasties, spanning roughly one thousand years, show several statistically significant features when rearranged in sidereal years. Most prominent is the peak of the number of records at the day 220, which coincides with the modern major periodic meteor shower, Perseids. The peak persists from Goryeo dynasty to Joseon dynasty, almost one thousand years. This implies that the comet 109P/Swift-Tuttle, the parent comet of Perseids, has been active for more than one thousand years without significant perturbation on its orbit. The change in the number of records for Perseids during Goryeo and Joseon dynasties suggests that the activity of Perseids has decreased over the same period. Another collection of peaks appear around the day 300, which can be associated with Orionids, north-Taurids, or Leonids. These peaks also suggest the activities of some of these major periodic meteor showers have increased over the last millennium.

We also find the evidence of seasonal variation of sporadic meteors in Korean records. According to Yrjölä & Jenniskens (1998), the activity of sporadic meteors is expected to peak at the Autumnal equinox and to be lowest at the Vernal equinox in the northern hemisphere. We confirm the seasonal variation of sporadic meteors from the records of Joseon dynasty, with the maximum number roughly 1.7 times the minimum. The least-square sinusoidal fit shows a phase shift of 15 days, toward winter, with respect to the simple geometrical expectation. We can think of two possible causes for the marginally detected phase difference. One factor could be the residual contamination by meteor showers. Another is the real innate seasonal variation of sporadic meteors. It is possible for the debris of each comet not to completely dissipate in a few years, resulting in excess of sporadic meteors near periodic meteor shower. In fact, there are many periodic meteor shower events in the winter, and the scattered meteoroids related with the meteor showers may have shifted the peak of the seasonal variation of sporadic meteors towards the winter.

The Chinese meteor shower and Japanese meteor records are also analyzed. The major features in the distribution of Chinese meteor shower records are quite consistent with those in Korean meteor records for the last one thousand years. Chinese records also suggest decrease in Perseids activity and increase in Orionids/north-Taurids/Leonids activity as the Korean records do. Japanese records also show Perseids feature and Orionids/north-Taurids-Leonids feature, although they are less prominent compared to Korean or Chinese records. One of the modern major periodic meteor shower, eta-Aquarids, which is one of the pair remnant of the comet Halley, dose not have any corresponding feature in Korean, Chinese, and Japanese records while another pair remnant of Halley, Orionids, appears somewhat strongly at the day 295. This may suggest eta-Aquarids has been active only recently.

Ahn et al. (2002) and Ahn (2003) have analyzed the meteor and meteor shower records of Goryeo dynasty. They have used the anomalistic year as the basic period of meteor showers. Even though the difference between the sidereal year (365.2564 days) and the anomalistic year (365.2596 days) is small, we point out that the sidereal year is the correct basic period. Ahn et al. (2002) and Ahn (2003) have analyzed the meteor and meteor shower records of Goryeo dynasty. They have used the anomalistic year as the basic period of meteor showers. Even though the difference between the sidereal year (365.2564 days) and anomalistic year (365.2596 days) is small, we point out that the sidereal year is the correct basic period. Ahn et al. (2002) and Ahn (2003)'s works were not too erroneous since the analysis were limited only for Goryeo records. But for longer period, sidereal year should be chosen as illustrated in Fig. 7. Fig. 7 shows how the Perseids peak looks when the sidereal and anomalistic years are used. We used 735 meteor records of Goryeo dynasty and 735 of Joseon dynasty. In Fig. 7a the Perseids peak is well-centered at the day 223. But the peak is skewed in Fig. 7b, which is due to the wrong choice of the basic period.

**(Figure 7)**

Babadzhanov (1994) has suggested five stages in the evolution of meteoroid stream. At its final stage, meteor stream becomes almost indistinguishable from the sporadic background. Evolution of the stream left by a comet depends on the condition of the comet. However, bright meteors decrease year by year after the comet passes. According to Wu (2002), Leonid showers would never be seen for more than about four consecutive years. Yeomans (1981) on the other hand has suggested that significant Leonid meteor showers were maintained for roughly 2500 days. We have looked for this kind of change in meteor activity in Korean records, wishing to estimate the evolution time of identified periodic meteor showers. For example, 231 records associated with Perseids are folded by using its period of 135 years, and distribution of the number of records as a function of phase is constructed. We have found no significant change of activity over cycles.

**Acknowledgements**

We thank M. McEachern and C. Lafon of the Smithsonian Astrophysical Observatory for kindly providing us copies of papers on historical meteor showers. This work is supported by Korea Science & Engineering Foundation through Astrophysical Research Center for the Structure and Evolution of the Cosmos.

**REFERENCES**

Ahn, S. H. 2003. Meteors and Showers a millennium ago. Mon. Not. R. Astron. Soc. 343, 1095-1100

Ahn, S. H., Bae, H, J., Cho, H. J., and Jung, S. W. 2002. Meteor showers of 10th to 14th century. Publications of the Korean Astronomical Society 17, 23-40

Ahn, Y. S., Lee, Y. B., Lee, Y. S., and Han, B. S. 2000. The Chronological Table during the Three Kingdoms period (三國時代年曆表). Korea Astron. Obs., Daejeon, Korea

Ahn, Y. S., Shim, K. J., Han, B. S., and Song, D. J. 2000. The Chronological Table during Joseon dynasty (朝鮮時代年曆表). Korea Astron. Obs., Daejeon, Korea

Babadzhanov, P. B. 1994. Asteroids and Their Meteor Showers. Yoshihide, K., Richard P.B., and Tomohiro, H. (Eds.), SFYHAF ASP Conference Series. 63, 168-185

Beijing observatory 1988. General Compilation of Chinese Ancient Astronomical Records (中國古代天象記錄總集). Jiangsu, China

Cha, C. S. 1983. The Official Histories of the Early Yi Dynasty. The Kyungpook Historical Review 6, 69-112

Chanjipcheong (Bureau of compilation) 1908. Jeungbomunheonbigo (增補文獻備考; Reference to the Old Books, enlarged with Supplements)

Chunchugwan (The Office for Annals Compilation) 1392-1863. Joseonwangjosillok (朝鮮王朝實錄; the Annals of the Joseon dynasty)

Han, B. S. 2001. Korean Calendar Conversion Table (韓國年曆大典). Yeungnam Univ. Press. Daegu, Korea

Hasegawa, I. 1980. Catalogue of ancient and naked-eye comets. Vistas Astron. 24, 59-102

Hasegawa, I. 1992. Historical Variation in the Meteor Flux as found in Chinese and Japanese Chronicles. Clest. Mech. Dynam. Astro. 54, 129-142

Hasegawa, I. 1998. Historical Variations in the Meteor Flux as Found in the History of the Koryo Dynasty. Yabushita, S., and Henrard, J. (Eds), Dynamics of Comets and Asteroids and their Role in the Earth History, Kluwer, Dordrecht/Boston/London

Imoto, S. and Hasegawa, I. 1958. Historical Records of Meteor Showers in China, Korea, and Japan. Smithson. Contrib. Astrophys 2, 131-144

Kanda, S. 1935. Japanese Historical Astronomical Records (日本天文史料). Tokyo, Japan

Kanda, S. 1936. General Inventory of Historical Astronomical Records

(日本天文史料綜覽). Tokyo, Japan

Kim, B. S. et al. 1145. Samguksagi (三國史記; the History of the Three Kingdoms).

Kim, J. S. et al. 1451. Goryeosa (高麗史; the History of the Goryeo dynasty).

Mason, J. W. 1995. The Leonid meteors and comet 55P/Tempel-Tuttle. JBAA 105, 219-235

Newton, H. A. 1864. The original accounts of the displays in former times of the November star-shower …(i), Amer. J. Sci. & Arts. 37, 377-389

Ohsaki, S. 1994. Japanese Historical Astronomical Records after 1600 (近世日本天文史料). Tokyo, Japan

Park, C. 1996. Analysis of Japanese Historical Solar Eclipse Records. Journal of Korean History of Science Society 18-2, 155-166

Park, C., and La, D. I. 1994. Confirmation and Historical Consequences of Astronomical Records in Samguksagi. Journal of Korean History of Science Society 16-2, 167-202

Rada, W. S., and Stephenson, F. R. 1992. A Catalogue of Meteor Showers in Medieval Arab Chronicles. QJRAS 33, 5-16

Shim, K. J., Ahn, Y. S., Han, B. S., Yang, H. J., and Song, D. J. 1999. The Chronological Table during Goryeo dynasty (高麗時代年曆表). Korea Astron. Obs. Daejeon, Korea

Yang, H. J., Park, C., and Park, M.-G. 1998. Evidence for the Solar cycle in the Sunspot and Aurora records of Goryer dynasty. Publications of the Korean Astronomical Society 13, 181-208

Yeomans, D. K. 1981. Comet Tempel-Tuttle and the Leonid meteors. Icarus 47, 492-499

Yrjölä, I., and Jenniskens, P. 1998. Meteor stream activity. A&A 330, 739-752

Williams, I. P. 1997. The Leonid meteor shower. Mon. Not. R. Astron. Soc. 292, L37-L40

Wu, G. J. 2002. Comet Tempel-Tuttle and the recent Leonid meteor shower. ChA&A 26, 40-48

Zhang, T. S. 1977. Ancient Chinese records of Meteor showers. Chinese Astronomy 1, 197-220

TABLE 1. Meteor, meteor shower, and meteorite records of Korea, China, and Japan divided into three periods.

| Country | Period | Interval (yr) | Records of | | |
|---|---|---|---|---|---|
| | | | meteor | meteor shower | meteorite |
| Korea | **The Three Kingdoms period** (−57 ~ 935) | 991 | 26 | 15[*] | 11 |
| | **Goryeo dynasty** (918 ~ 1392) | 475 | 731 | 9 | 38 |
| | **Joseon dynasty** (1392 ~ 1910) | 519 | 3104 | 7 | 5 |
| China | −645 ~ 918 | 1565 | 368 | 35 | 60 |
| | 919 ~ 1391 | 473 | 1786 | 16 | 25 |
| | 1392 ~ 1911 | 520 | 3521 | 225 | 299 |
| Japan | 636 ~ 917 | 282 | 73 | 5 | 15 |
| | 918 ~ 1391 | 474 | 78 | 5 | 2 |
| | 1392 ~ 1867 | 476 | 280 | 3 | 14 |

---

[*] Two of 15 records do not have any month or date information

TABLE 2. Historical Korean meteor shower records compiled from the three official Korean history books (*Samguksagi, Goryeosa*, and *Joseonwangjosillok*) and *Jeungbomunheonbigo*.

| Date of Observation[1] ( Y  M  D )[2] | J D[3] | Day[4] | Associated Meteor shower | Estimated date of shower[5] | Period | Ref.[6] | Note[7] |
|---|---|---|---|---|---|---|---|
| 104 2/3 – | 1759102 | 82 | | | | S, M | 1 |
| 454 – – | – | – | | | | S, M | 1 |
| 532 8 28 | 1915611 | 261 | | | | S, M | 1 |
| 566 6/7 – | 1927955 | 190 | | | The Three | M | 1 |
| 581 3 20 | 1933347 | 100 | | | | S, M | 1 |
| 586 5/6 – | 1935252 | 178 | | | Kingdoms | S, M | 1 |
| 643 11 1 | 1956218 | 325 | | | | S, M | 1 |
| 647 9/10 – | 1957636 | 282 | | | Period | S, M | 1 |
| 684 11/12 – | 1971220 | 351 | | | | S, M | 1 |
| 706 4/5 – | 1979045 | 141 | | | | S, M | 1 |
| 718 10/11 – | 1983622 | 334 | | | | S, M | 1 |
| 765 1 7 | 2000481 | 26 | | | | S | 1 |
| 801 10/11– | 2013921 | 317 | Leonids[8] ? | 801 10 11[B] | | S, M | 1 |
| 848 – –[9] | – | – | | | | M | 1 |
| 905 3/4 – | 2051691 | 101 | | | | S, M | 1 |
| 1042 7 25 | 2101854 | 223 | Perseids | 1042 7 25[C] | | G, M | 1 |
| 1095 7 25 | 2121212 | 223 | Perseids | 1095 7 25[C] | | G, M | 3 |
| 1103 9 7 | 2124178 | 267 | | | Goryeo | G, M | 3 |
| 1106 7 27 | 2125232 | 225 | Perseids | 1106 7 25[C] | | G, M | 3 |
| 1111 10 2 | 2127125 | 292 | Leonids | 1111 10 11[B] | Dynasty | G, M | 3 |
| 1136 4 3 | 2136075 | 110 | | | | G, M | 1 |
| 1178 9 17 | 2151582 | 277 | | | Period | G | 2 |
| 1179 4 17 | 2151794 | 123 | η Aquarids | 1179 4 20[C] | | G, M | 3 |
| 1363 6 9[10] | 2219053 | 175 | | | | G, M | 1 |
| 1532 10 24 | 2280918 | 312 | Leonids | 1532 10 24.9 [A] | | J, M | 1 |

[1] Julian calendar for dates prior to Oct. 4, 1582, Gregorian calendar after Oct. 15, 1582.
[2] Records with no month or day information are noted by "–".
[3] The record with no day information is set to be the fifteenth day of the recorded lunar month.
[4] The counting day within each sidereal year. Jan. 1, 2000 is set to be the day 1 of the reference sidereal year.
[5] A is from Mason (1995), B is calculated by interpolating Mason's dates, and C is estimated by using the sidereal period of each periodic shower.
[6] S means *Samguksagi*, G is *Goryeosa*, J is *Joseonwangjosillok*, and M is *Jeungbomunheonbigo*.
[7] 1: records mentioned in the catalogue of Imoto & Hasegawa (1958). 2: date of observation different from that of Imoto & Hasegawa's catalog (the source book they used has a typo in date information). 3: not listed in the Imoto & Hasegawa's catalog.
[8] This record was not listed in Table 1 of Mason (1995).
[9] This record has only seasonal information, "fall".
[10] This record is dubious as a meteor shower phenomenon.

| | | | | | | | |
|---|---|---|---|---|---|---|---|
| 1533 10 24 | 2281283 | 312 | Leonids | 1533 10 25.0 [A] | Joseon | J | 1 |
| 1548 8 24 | 2286701 | 251 | | | | J, M | 1 |
| 1560 8 24 | 2291084 | 251 | | | Dynasty | J | 1 |
| 1566 10 26 | 2293338 | 313 | Leonids | 1566 10 25.9 [A] | | J | 1 |
| 1602 11 12 | 2306494 | 320 | Leonids | 1602 11 6.9 [A] | Period | J | 1 |
| 1625 11 6 | 2314889 | 314 | Leonids | 1625 11 5.9 [A] | | J | 1 |

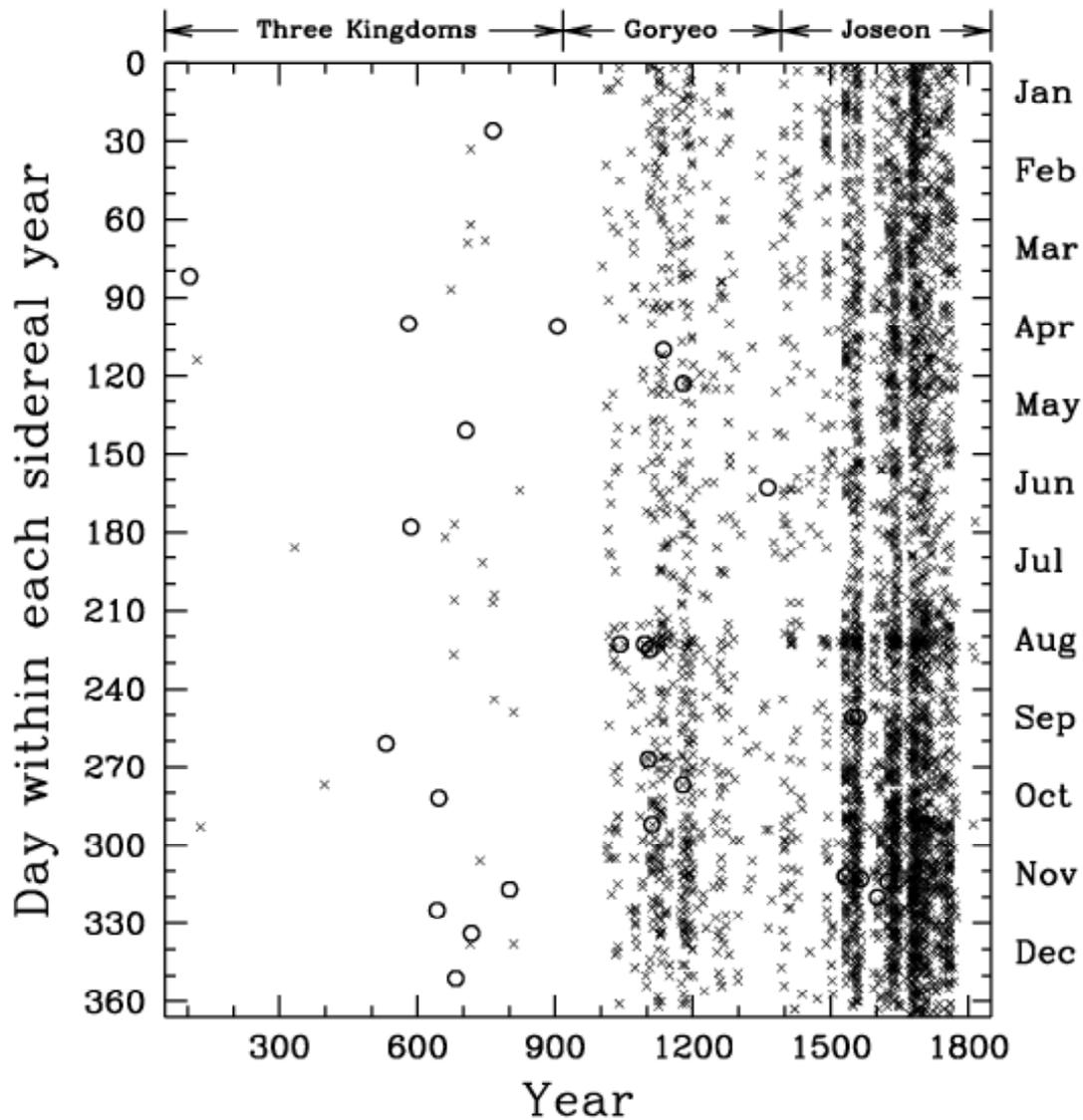

FIG. 1.    Distribution of 3860 historical Korean meteor and 28 meteor shower records from 57 B.C. to A.D. 1910. Records are compiled from the three official Korean history books: *Samguksagi* ( 三國史記, the History of the Three Kingdoms), *Goryeosa* (高麗史, the History of the Goryeo dynasty), and *Joseonwangjosillok* (朝鮮王朝實錄, the Annals of the Joseon dynasty). The circles are meteor showers, and crosses are meteors. All dates recorded in history books are converted to days in sidereal year. January 1, A.D. 2000 is set to be the day 1 of the reference sidereal year.

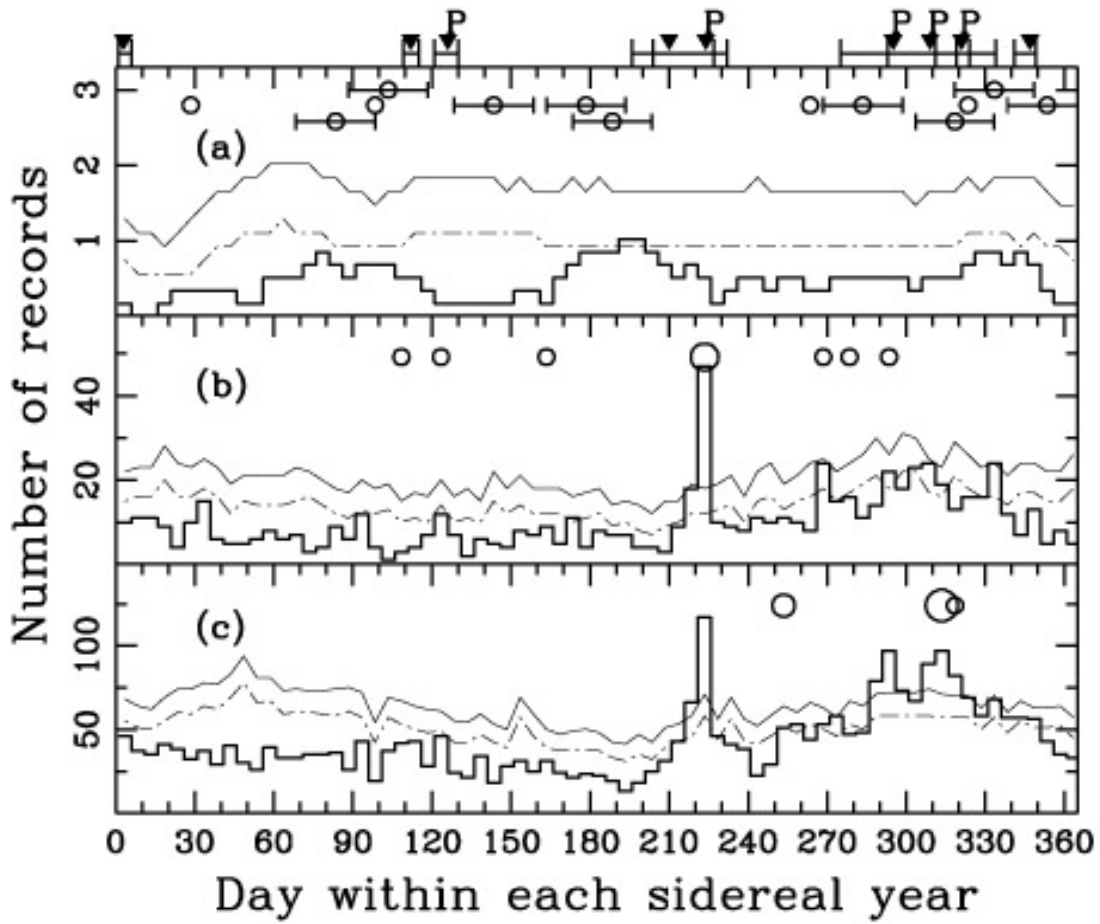

FIG. 2. Distributions of historical Korean meteor and meteor shower records divided into three periods. The bin size of the histograms is 5 days. Meteor shower records are shown as open circles in the upper part of each panel and the size of each circle represents the number of meteor shower events. (a) 38 records including 12 meteor showers during the Three Kingdoms period (57 B.C.~A.D. 935). The meteor shower records without day information have one-month uncertainty. (b) 740 records including 9 meteor showers during Goryeo dynasty (A.D. 918~1392). (c) 3111 records including 7 meteor showers during Joseon dynasty (A.D. 1392~1910). January 1, A.D. 2000 is set to be the day 1 of the reference sidereal year, and each date is converted to day within a sidereal year. The horizontal error-bars and triangles above the top panel indicate the durations and maximum dates of well-known meteor shower events as of A.D.f2000, respectively. And bold character P's above the triangles represent periodic meteor showers. The lines in each box are confidence levels of 90% (1.7 σ; dot-dashed line) and 99.9% (3 σ; solid line) estimated from Monte Carlo simulation. Monte Carlo samples were randomly generated to have the same number of data from distribution of astronomical records which can represent seasonal observational bias during the period of each dynasty.

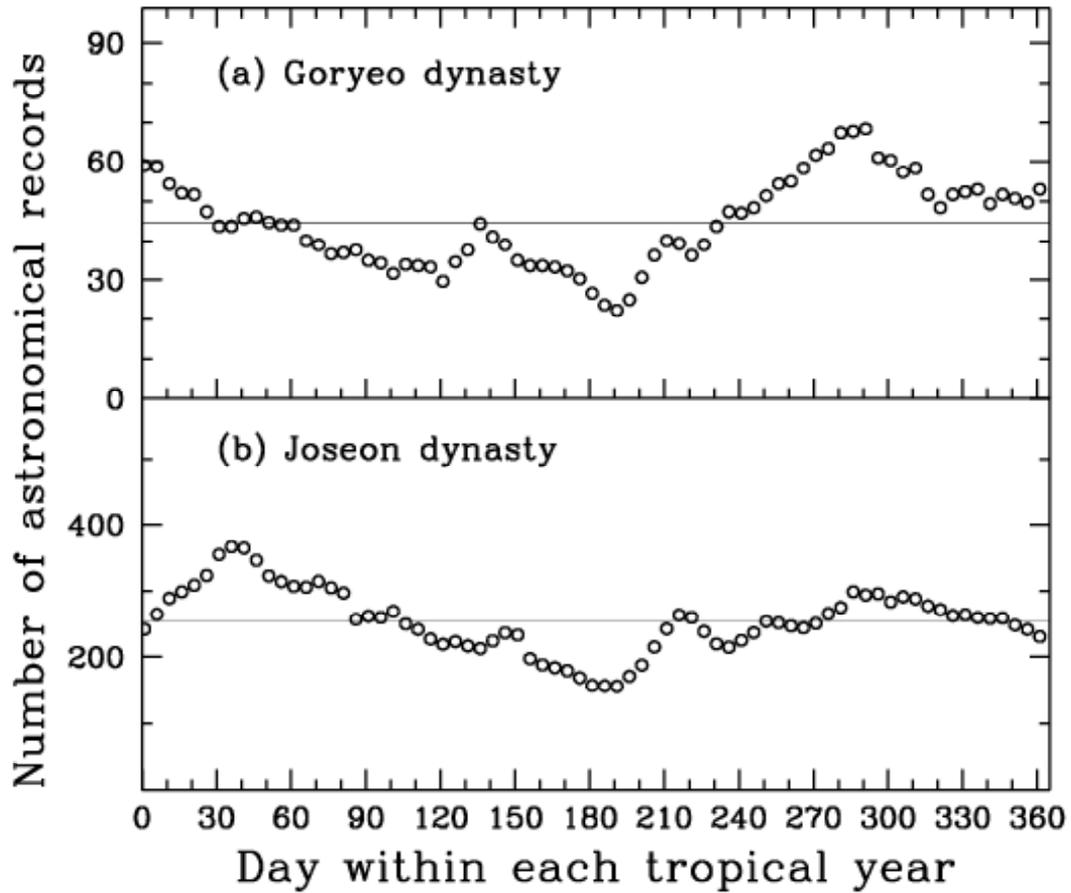

FIG. 3.　Seasonal observational bias, i.e. number distributions of certain class of astronomical records during Goryeo (A.D. 918-1392) and Joseon (A.D. 1392-1910) dynasties. The tropical year is used as the basic period. Planetary observation records are mainly used. Since the number of planetary observation of Goryeo dynasty is not large enough to clearly see the seasonal observational bias, lunar observation records are added for Goryeo dynasty. The numbers of selected records used for Goryeo and Joseon dynasties are 3534 and 18697, respectively.

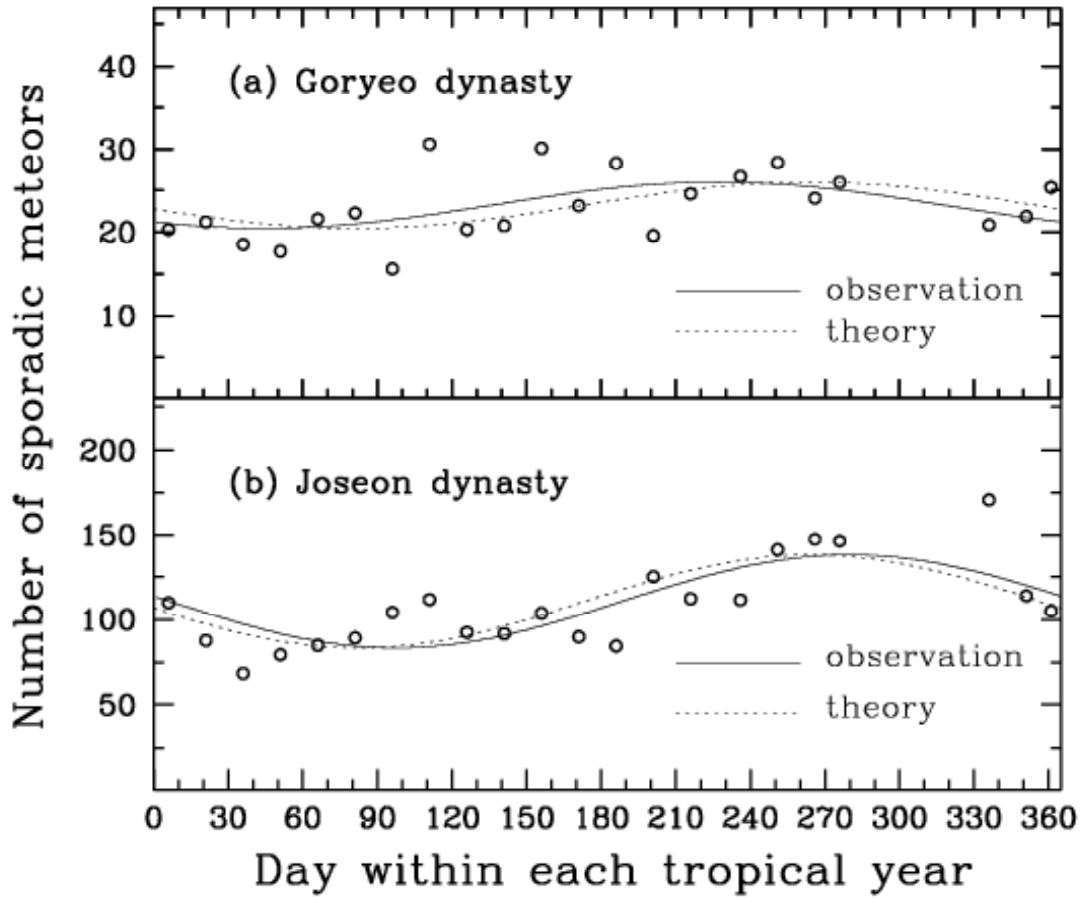

FIG. 4. Seasonal variations of sporadic meteor records during Goryeo (A.D. 918-1392) and Joseon (A.D. 1392-1910) dynasties. The seasonal distributions of records are divided by the normalized seasonal observational biases (Fig. 3) to produce the inherent seasonal variations (circles) of sporadic meteors during Goryeo and Joseon dynasties. The solid curve is a best fit model (Eq. 3) with three free parameters, the mean, amplitude of variation, and the phase shift. The dotted curve is a theoretical prediction with zero phase shift. The maximum number of sporadic meteors in a year is roughly 1.7 times the minimum during Joseon dynasty.

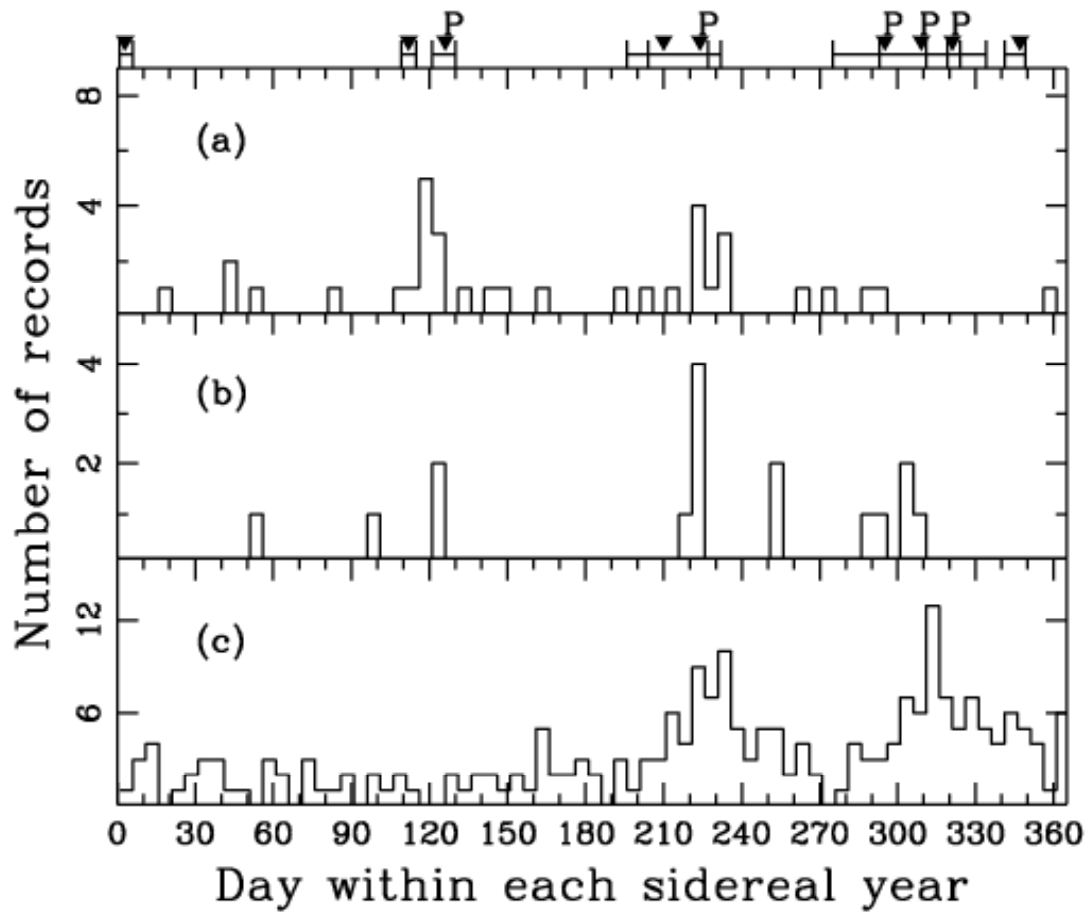

FIG. 5. Distributions of historical Chinese meteor shower records. All records are divided into the same three periods used for Korean records, to facilitate comparison with Korean records: (a) before A.D. 918, (b) from A.D. 918 to A.D. 1392, and (c) from A.D. 1393 to A.D. 1911. Each period contains 35, 16, and 225 meteor shower records, respectively. All treatments of data and definitions of symbols are the same as in Fig. 2.

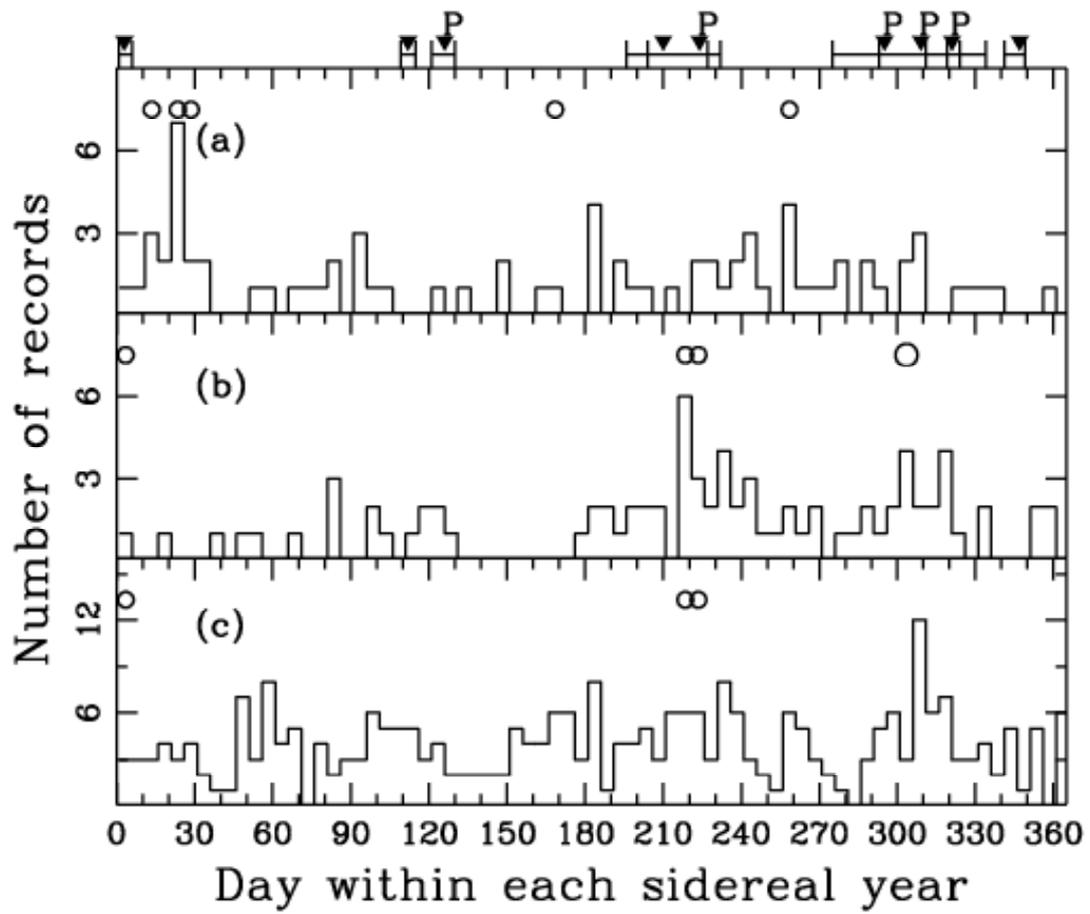

FIG. 6.    Distributions of historical Japanese meteor and meteor shower records. All records are divided into the same three periods. All treatments of data and definitions of symbols are the same as in Fig. 2.

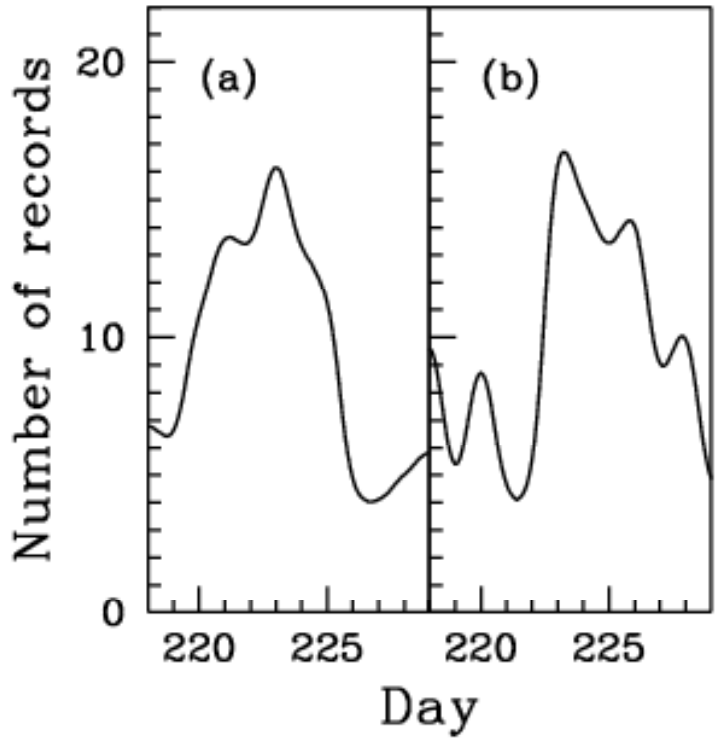

FIG. 7.    Distributions of meteor records near the Perseids peak during Goryeo and Joseon dynasties. The sidereal (365.2564 days) and anomalistic (365.2596 days) years are used as the basic meteor shower period in panel (a) and (b), respectively. The bin size is one day.